\input harvmac
\input psfig
\newcount\figno
\figno=0
\def\fig#1#2#3{
\par\begingroup\parindent=0pt\leftskip=1cm\rightskip=1cm\parindent=0pt
\global\advance\figno by 1
\midinsert
\epsfxsize=#3
\centerline{\epsfbox{#2}}
\vskip 12pt
{\bf Fig. \the\figno:} #1\par
\endinsert\endgroup\par
}
\def\figlabel#1{\xdef#1{\the\figno}}
\def\encadremath#1{\vbox{\hrule\hbox{\vrule\kern8pt\vbox{\kern8pt
\hbox{$\displaystyle #1$}\kern8pt}
\kern8pt\vrule}\hrule}}
\def\underarrow#1{\vbox{\ialign{##\crcr$\hfil\displaystyle
 {#1}\hfil$\crcr\noalign{\kern1pt\nointerlineskip}$\longrightarrow$\crcr}}}
%
\overfullrule=0pt
\def\hat{\widehat}
%
\def\tilde{\widetilde}
\def\bar{\overline}
\def\Z{{\bf Z}}

\def\S{{\bf S}}

\def\R{{\bf R}}

\font\zfont = cmss10 
\font\litfont = cmr6

\def\bigone{\hbox{1\kern -.23em {\rm l}}}
\def\ZZ{\hbox{\zfont Z\kern-.4emZ}}
\def\half{{\litfont {1 \over 2}}}

\def\N{{\cal N}}

\Title{hep-th/9812012, IASSNS-HEP-98-96}
{\vbox{\centerline{AdS/CFT Correspondence}
\bigskip
\centerline{ And Topological Field  Theory    }                }}
\smallskip
\centerline{Edward Witten}
\smallskip
\centerline{\it School of Natural Sciences, Institute for Advanced Study}
\centerline{\it Olden Lane, Princeton, NJ 08540, USA}\bigskip

\medskip

\noindent
In $\N=4$ super Yang-Mills theory on a four-manifold $M$, one can specify a 
discrete magnetic flux valued in $H^2(M,\Z_N)$.  This flux  is encoded in the 
AdS/CFT correspondence in terms of a five-dimensional topological field theory
with Chern-Simons action.  A similar topological field theory in seven 
dimensions governs the space of ``conformal blocks'' of the six-dimensional 
$(0,2)$ conformal field theory.
\Date{December, 1998}
\newsec{Introduction}
The AdS/CFT correspondence \ref\malda{J. Maldacena, ``The Large $N$
Limit Of Superconformal Field Theories And Supergravity,''
hep-th/9711200.} relates
 $\N=4$ super Yang-Mills theory on $\S^4$, with gauge
group $SU(N)$, to Type IIB superstring theory on $AdS_5\times \S^5$,
with $N$ units of five-form flux on $\S^5$.
The correspondence expresses gauge theory correlation functions
\nref\kleb{S. S. Gubser, I. R. Klebanov, and A. M. Polyakov,
``Gauge Theory Correlators From Non-Critical String Theory,''
hep-th/9802109.}
\nref\witten{E. Witten, ``Anti de Sitter Space And Holography,''
hep-th/9803002.}
in terms of the dependence of the string theory on the behavior
near the conformal boundary of $AdS_5$ \refs{\kleb,\witten}.

What happens if we replace $\S^4$ with a more general four-dimensional
spin manifold $M$?  ($M$ must be spin since the $\N=4$ theory contains
spinor fields.  In addition, $M$ must be endowed with a metric that
-- after a suitable conformal rescaling -- has positive scalar curvature,
or the $\N=4$ theory is unstable.\foot{The instability arises because
of possible runaway behavior of the massless scalars $\phi$ of the theory.
Positivity of the scalar curvature $R$ of $M$
suppresses the instability because of the $R\,\,\tr \,\phi^2$ coupling
required by conformal invariance.})
The behavior of the gauge
theory on a four-manifold $M$ is believed to be
described, roughly speaking, 
in terms of the behavior of
the string theory on $X\times \S^5$, where $X$ is a negatively curved
Einstein manifold with $M$ as conformal boundary, and one must sum over
all possible choices of $X$.  This description is somewhat rough
since, in general, one might need to include branes or stringy singularities
on $X\times \S^5$ or more general topologies that are not simply the product
of $\S^5$ with some five-manifold.  The general prescription is really
that one sums over Type IIB spacetimes that near infinity look like
$X\times \S^5$, with $X$ a five-manifold of conformal boundary $M$.

The gauge theory on $M$ that is described in this correspondence has a gauge
group that is locally $SU(N)$, rather than $U(N)$.  Many arguments show
this, beginning with the
fact that in $\N=4$ super Yang-Mills
theory of $U(N)$, the $U(1)$ fields would be free, while in
string theory on $AdS_5\times \S^5$, everything couples to gravity and
no field is free.  As the gauge group is $SU(N)$, its center is $\Z_N$.

One consequence of the fact that the gauge group is $SU(N)$ rather
than $U(N)$ is that it is possible to make a baryon vertex linking
$N$ external quarks.  This can be constructed using a wrapped fivebrane
\ref\bary{E. Witten, ``Baryons And Branes In Anti de Sitter Space,''
hep-th/9805112.} and can also be understood
\ref\googuri{D. J. Gross and H. Ooguri, ``Aspects
of Large $N$ Gauge Theory Dynamics As Seen By String Theory,''
hep-th/9805129.} via a certain Chern-Simons interaction
that we will describe shortly.

Our interest in the present paper will, however, be in properties that
depend on the global structure of $M$.
In $\N=4$ super Yang-Mills theory, all fields are in the adjoint representation
and hence, as a local gauge transformation,
 the center of the gauge group acts trivially.
It follows that locally, the gauge group could be $SU(N)/\Z_N$.  This
has two important consequences that depend on the  topology of $M$:

(1) If $H^1(M,\Z_N)\not= 0$, it is possible to consider gauge transformations
on $M$ that are not single-valued in $SU(N)$ but are single-valued in
$SU(N)/\Z_N$.   For example, if $M=\S^1\times Y$, one can consider
a gauge transformation that is constant in the $Y$ direction and whose
restriction to the $\S^1$ direction determines any element of
$\pi_1(SU(N)/\Z_N)=\Z_N$.  This gives an important group $F\cong\Z_N$
of global symmetries of the thermal physics on $Y$.
More generally, $F=H^1(M,\Z_N)$ classifies $SU(N)/\Z_N$ gauge transformations
that cannot be lifted to $SU(N)$.

(2) If $H^2(M,\Z_N)\not= 0$, it is possible to consider $SU(N)/\Z_N$
bundles with ``discrete magnetic flux'' \ref\thooft{G. 't Hooft,
``On The Phase Transition Towards Permanent Quark Confinement,''
Nucl. Phys. {\bf B138} (1978) 1, ``A Property Of Electric
And Magnetic Flux In Nonabelian Gauge Theories,'' {\bf B153} (1979) 141.}.
In fact, $SU(N)/\Z_N$ bundles on $M$ are classified topologically
by specifying the instanton number and also a characteristic class $w$ that 
takes values in $K=H^2(M,\Z_N)$.  (For example, if
$N=2$, we have $SU(2)/\Z_2=SO(3)$, and $w$ coincides with the second 
Stieffel-Whitney class of the gauge bundle.)

\nref\witto{E. Witten, ``Fivebrane Effective Action In
$M$-Theory,'' J. Geom. Phys. {\bf 22} (1997) 103.}
How to see in the AdS/CFT correspondence the thermal symmetry $F$ has
been discussed elsewhere \ref\ahaw{O. Aharony and E. Witten, ``Anti
de Sitter Space And The Center Of The Gauge Group,'' hep-th/9807205.}.
For our present purposes, we need to recall just one point from
that discussion.  Type IIB superstring theory has two two-form fields
$B_{NS}$ and $B_{RR}$.  In compactification on $X\times \S^5$ (with $N$ units
of five-form flux on $\S^5$), one gets a low energy effective action
on $X$ that contains a Chern-Simons term
\eqn\huxx{L_{CS}=NI_{CS},}
where 
\eqn\juxx{I_{CS}=-{i\over 2\pi}\int_X B_{RR}\wedge dB_{NS}}
is the basic Chern-Simons invariant of the two $B$-fields.
(The action $L_{CS}$ 
is important in one approach to the baryon vertex \googuri.)
Though the integrand in $I_{CS}$ is not gauge-invariant, $I_{CS}$ is
gauge-invariant mod $2\pi i$ on a closed five-manifold $X$.  It can
be written more invariantly as follows: if $X$ is
the boundary of a six-manifold $Y$ over which the two $B$-fields extend, and
we write the field strength of a $B$-field as $H=dB$,
then one can write $I_{CS}$ in a manifestly gauge-invariant way as
\eqn\zuxx{I_{CS}=-{i\over 2\pi}\int_Y H_{RR}\wedge H_{NS}.}
More generally, if $Y$ does not exist, a more subtle approach is needed
to define $I_{CS}$.  For a general Chern-Simons interaction, one 
can follow a slightly abstract approach 
explained in section 2.10
and the end of the introduction to section 4 in \witto.
For the particular Chern-Simons theory that we are discussing
here, one can define $I_{CS}$ as a cup product in Cheeger-Simons
cohomology.\foot{In the language of Cheeger and Simons
\ref\cheeger{J. Cheeger and
J. Simons, ``Differential Characters And Geometric Invariants,''
in {\it Geometry and Topology}, ed. J. Alexander and A. Harer,
Lecture Notes in Mathematics vol. 1167 (Springer-Verlag, 1985).},
a $B$-field on a manifold $X$ is an element of $\hat H^2(X,U(1))$,
the group of differential two-characters on $X$ with values in $U(1)$.
For two elements $B_{RR},B_{NS}\in \hat H^2(X,U(1))$,
Cheeger and Simons describe in Theorem 1.11
a product $B_{RR}*B_{NS}\in \hat H^5(X,U(1))$.
For $X$ of dimension five, $\hat H^5(X,U(1))=U(1)$, and we set
$\exp{\left(-I_{CS}(B_{RR},B_{NS})\right)}=B_{RR}*B_{NS}$.
Equation 1.15 in \cheeger\ asserts that if $B_{RR}$ is topologically
trivial and hence can be defined by an ordinary two-form,
$I_{CS}$ can be computed by the integral written in \juxx:
$I_{CS}=-(i/2\pi)\int_X B_{RR}\wedge H_{NS}$.  Of course, there is
a similar formula $I_{CS}=(i/2\pi)\int_X H_{RR}\wedge B_{NS}$ if
$B_{NS}$ is topologically trivial.  It can also be proved that
$B_{RR}*B_{NS}$ can be defined by the formula in the text if $X$ is the
boundary of an appropriate six-manifold $Y$.
The usefulness of interpreting
the action in terms of differential characters was explained to me
by M. Hopkins, who also pointed out facts that we will use in
section 3.4.}

A theory of the two $B$-fields with Lagrangian precisely \huxx\ is
a simple but subtle topological field theory, of a sort first
considered in \ref\aschwarz{A. Schwarz, ``The Partition Function Of
Degenerate Quadratic Functional And Ray-Singer Invariants,''
Lett. Math. Phys. {\bf 2} (1978) 247.}.
The actual low energy
effective action that arises in compactification of Type IIB superstring
theory on $\S^5$ has many couplings beyond the Chern-Simons interaction;
the additional  couplings have however a larger number of derivatives and
so are irrelevant for certain questions.

The goal of the present paper is to understand  
how to encode in terms of the AdS/CFT correspondence the
dependence of the gauge theory on the discrete magnetic flux $w$.
As we will see, the topological field theory with Lagrangian \huxx\ plays
a starring role in the analysis.  Roughly speaking, what in the gauge
theory description appears as the discrete magnetic flux $w$ appears
in the gravitational description on $X\times \S^5$ as a quantum state
of this topological field theory.  We state a precise conjecture in
section 2 and carry out some computations supporting it in section 3.
Then in section 4, we make an extension to the $(2,0)$ superconformal
field theory in six dimensions, and in section 5 we discuss 't Hooft and
Wilson loops in the four-dimensional $\N=4$ theory.

Since we will be developing a fairly elaborate theory to understand
in terms of gravity the dependence of the gauge theory on the discrete
magnetic flux, it is
reasonable to ask what examples are known to which this theory can
be applied.  Actually, as we noted at the outset, there are relatively
few $M$'s for which the AdS/CFT correspondence is expected to work ($M$ must
be spin and with positive scalar curvature),
and there are presently very few such examples for which
anything is known about the possible five-manifolds $X$ with conformal
boundary $M$.  One simple but important example, in which one can
verify the importance of summing over different choices of $X$, is the
case $M=\S^1\times \S^3$.  There are \ref\hawking{S. W. Hawking and D.
Page, ``Thermodynamics Of Black Holes In Anti de Sitter Space,''
Commun. Math. Phys. {\bf 87} (1983) 577.}
two known choices of $X$, namely $X_1=\S^1\times B_4$ and $X_2=B_2\times \S^3$
(here $B_n$ denotes an $n$-dimensional ball); many important properties
of Yang-Mills theory at nonzero temperature are reflected in the
behavior of these two $X$'s \ref\withermal{E. Witten,  ``Anti-de Sitter
Space, Thermal Phase Transition, And Confinement In Gauge Theories,''
hep-th/9803131.}.

This example  has $H^2(M,\Z_N)=0$ and so does not serve
as a good illustration of the issues explored in the present paper.  
But one can
readily modify it to give an example with nontrivial discrete magnetic
fluxes.  Identify $\S^3$ with the $SU(2)$ manifold and let
$H$ be a discrete subgroup of $SU(2)$, acting on the right.
  Set $M_H=\S^1\times \S^3/H$.  $M_H$ is the boundary
of $X_{i,H}=X_i/H$.  For suitable $H$, $H^2(M_H,\Z_N)\not= 0$, so this
gives a simple example to which our theory can be applied, showing
in particular that the theory is nonvacuous.

This example actually has the following interesting property.
$X_{2,H}$ has orbifold singularities (because $H$ acts freely on $\S^3$
but not on $B_4$).  The orbifold singularities are harmless in string theory
and will be resolved as one varies the metric on $M_H$.  The  manifold
 $X_{2,H}$ will  lose and acquire singularities and undergo monodromies
 as the metric on $M_H$ is varied.

Another example is $M=\S^2\times \S^2$, where some $X$'s have been
constructed in \ref\bohm{Christoph B\"ohm, ``Noncompact Cohomogeneity
One Einstein Manifolds,'' preprint.} and investigated independently
in \ref\muk{M. Rangamani, unpublished.}.  This example has the
property that $H^2(M,\Z)$ is not a torsion group.

\bigskip\noindent{\it Some Technical Details}

We conclude this introduction by summarizing a few useful details.
In this paper, the symbol $b_i$ will denote the $i^{th}$ Betti number
of the four-manifold $M$.  We also write the order of the finite group
$H^i(M,\Z_N)$ as $N^{\tilde b_i}$. (In general the $\tilde b_i$
are not integers.)   If there is no torsion in the integral
cohomology of $M$, then $b_i=\tilde b_i$.  Using Poincar\'e and Pontryagin
duality, it is possible to prove that $b_{4-i}=b_i$ and likewise
$\tilde b_{4-i}=\tilde b_i$.  
The last statement arises because there is a nondegenerate pairing
$H^i(M,\Z_N)\times H^{4-i}(M,\Z_N)\to \Z_N$ (given by the cup product); we write
the product of $x\in H^i(M,\Z_N)$ with $y\in H^{4-i}(M,\Z_N)$ as
$x\cdot y$.  Nondegeneracy of the pairing implies that 
these groups are of the same order (and in fact
are isomorphic as abelian groups).
One also has $b_0=\tilde b_0=1$.
We write $\chi $ and $\sigma$ for the Euler characteristic and signature
of $M$; we recall the definition $\chi=\sum_{i=1}^4(-1)^ib_i$.
Using  the long exact sequence of
cohomology groups derived from the short exact sequence of groups
$0\to \Z\underarrow{N} \Z\to \Z_N\to 0$ (the first map is multiplication
by $N$ and the second is reduction modulo $N$), one can prove that in fact
\eqn\kurro{\sum_{i=0}^4(-1)^ib_i=\sum_{i=0}^4(-1)^i{\tilde b_i}.}
Hence $b_1-\half b_2=\tilde b_1-\half \tilde b_2$, a fact we will use later.

\newsec{Role Of The Topological Field Theory}

\subsec{The Problem}

As a prelude to our main subject, let us ask: In studying the $\N=4$
super Yang-Mills theory on a four-manifold $M$ via the AdS/CFT correspondence,
do we expect to see Montonen-Olive
$S$-duality?  The answer is, ``Not in a naive way,''
since $SU(N)$ is not a self-dual group.  For example, if
\eqn\deftau{\tau={4\pi i\over g^2}+{\theta\over 2\pi}}
is the coupling parameter of the theory, then under $\tau\to -1/\tau$
we expect $SU(N)$ to be exchanged with $SU(N)/\Z_N$.

\nref\vafawitten{C. Vafa and E. Witten, ``A Strong Coupling Test Of
$S$-Duality,'' Nucl. Phys. {\bf B431} (1994) 3.}
The $SU(N)$ and $SU(N)/\Z_N$ theories are equivalent locally,
and they are essentially equivalent on a four-manifold such as $M=\S^4$
or $\S^3\times \S^1$ whose second cohomology group vanishes.
\foot{On such a manifold, the $SU(N)$ and $SU(N)/\Z_N$ theories
have the same correlation functions; their partition functions differ
by an elementary factor described in eqn. (3.17) of \vafawitten.  
This factor arises because the groups of $SU(N)$ and $SU(N)/\Z_N$ gauge
transformations differ slightly; the center of $SU(N)$ consists of gauge
transformations in $SU(N)$ that are not considered in $SU(N)/\Z_N$,
while  $F=H^1(M,\Z_N)$ classifies the classes of global $SU(N)/\Z_N$ gauge
transformations that cannot be lifted to $SU(N)$.  With
$N^{\tilde b_1}$  the order of the finite group $F$,
the group of $SU(N)$ gauge transformations has volume
$N^{1-\tilde b_1}$ times that for $SU(N)/\Z_N$, and hence
the $SU(N)$ partition function is $N^{-1+\tilde
b_1}$ times that for $SU(N)/\Z_N$,  a fact which is incorporated
in the next equation in the text.
If, as assumed in \vafawitten, there is no torsion in the cohomology of
$M$, then $\tilde b_1=b_1$, and the factor becomes $N^{-1+b_1}$,
as written in \vafawitten.}
These two theories really differ in an interesting way when $H^2(M,\Z_N)
\not= 0$.  The reason is that $SU(N)$ bundles on $M$ are classified
just by the instanton number, but $SU(N)/\Z_N$ bundles are classified
by an additional topological invariant which is the ``discrete magnetic
flux'' $w\in H^2(M,\Z_N)$ mentioned in the introduction.  The $SU(N)$ theory
has a partition function $Z(\tau)$ (we suppress $M$ from the notation
when this is likely to cause no confusion), but the $SU(N)/\Z_N$ theory
has a family of partition functions $Z_w(\tau)$, one for each
$w\in H^2(M,\Z_N)$.  The relation between the two is that the $SU(N)$
partition function is obtained from the $SU(N)/\Z_N$ partition function
by setting $w=0$, up to an elementary factor (mentioned in the footnote):
\eqn\relpar{Z_{SU(N)}(\tau)=N^{-1+\tilde b_1} Z_0(\tau).}

The $SU(N)$ theory is thus obtained by setting
$w=0$, but in the $SU(N)/\Z_N$ theory,
one sums over all $w$.  Since these operations are supposed to be
exchanged under $\tau\to -1/\tau$, clearly the contribution with
a given $w$ cannot be $SL(2,\Z)$-invariant.  

Generalizing ideas of 't Hooft \thooft, it has been argued \vafawitten\ that
the $Z_w(\tau)$ transform as a unitary representation of $SL(2,\Z)$. 
To describe this representation, one must use the fact that the cup
product gives a natural pairing
$H^2(M,\Z_N)\times H^2(M,\Z_N)\to H^4(M,\Z_N)=\Z_N$, as mentioned
at the end of the introduction. 
The most interesting part of the action of $SL(2,\Z)$ is the behavior
under $\tau\to -1/\tau$, which according to \vafawitten\ is a sort of
discrete Fourier transform:
\eqn\fourtran{Z_v(-1/\tau)= N^{-\tilde b_2/2}\left({\tau\over i}\right)^{W/2}
\left({\bar\tau\over -i}\right)^{\bar W/2}
\sum_{w\in H^2(M,\Z_N)}
\exp(2\pi i v\cdot w/N) \,Z_w(\tau).}
The intuitive idea of this formula is that $\tau\to -1/\tau$
exchanges magnetic and electric flux, but the discrete electric flux
is defined by a Fourier transform with respect to the magnetic variable.
The factor of $N^{-\tilde b_2/2}$ is needed for $S^2=1$, since
 the order of the finite group $K=H^2(M,\Z_N)$ is $N^{\tilde b_2}$.
(In \vafawitten, the cohomology was assumed to be torsion-free, and this
factor reduced to $N^{b_2}$.)
 The modular weights $W$ and $\bar W$ are expected to be linear functions
of $\chi$ and $\sigma$ (the Euler characteristic and signature of $M$).
They cannot be determined just from gauge theory, as they can be
modified by adding gravitational couplings of the general form $f(\tau,\bar 
\tau)RR$  ($R$ being the Riemann tensor of $M$); 
to predict the modular weights
that are observed in computing the $Z_v(\tau)$ using the AdS/CFT correspondence,
one would need to know precisely which such couplings are determined
by this correspondence.  (In \vafawitten, $\bar W$ was set to zero,
since a twisted version of the theory with a holomorphically varying partition
function was considered.)

In addition to \fourtran, the $SL(2,\Z)$ transformation law of the $Z_v(\tau)$
is specified by describing the behavior under $\tau\to \tau+1$.  This is
described in eqn. (3.14) of \vafawitten\ and is determined by the fact
that the instanton number of an $SU(N)/\Z_N$ bundle is not an integer
but is of the form (if $M$ is a spin manifold)
\eqn\jukk{{v^2\over 2N}\,\,{\rm modulo}\,\,\Z,}
as in equation (3.13) of \vafawitten.   (A term $v^2/2$, which is integral
if $M$ is spin, has been omitted.)
The transformation $\tau\to \tau+1$ amounts to $\theta\to \theta+2\pi$
in gauge theory, so the transformation law is
\eqn\mukk{Z_v(\tau+1)=\exp\left(2\pi i(v^2/2N-s)\right)Z_v(\tau),}
where $s$ is a constant that reflects the fact that the gravitational
couplings $f(\tau,\bar\tau)RR$ may not be invariant under $\tau\to\tau+1$.

It follows from \relpar\ and \fourtran\ that, essentially as in eqn. (3.18) of
\vafawitten, the $SU(N)$ and $SU(N)/\Z_N$  partition functions are
related by
\eqn\relpar{Z_{SU(N)}(-1/\tau)=N^{-\chi/2}\left({\tau\over i}\right)^{W/2}
\left({\bar \tau\over -i}\right)^{\bar W/2}Z_{SU(N)/\Z_N}(\tau).}
This is essentially the Montonen-Olive formula, saying that $\tau\to -1/\tau$
exchanges $SU(N)$ and $SU(N)/\Z_N$; the prefactors reflect gravitational
couplings not present in the original Montonen-Olive formulation on
flat $\R^4$.\foot{The derivation of \relpar\ in \vafawitten\ assumed no torsion
in $H^2(M,\Z_N)$ and  gave for the first factor $N^{-1+b_1-b_2/2}=N^{-\chi/2}$.
More generally, including the torsion, one gets $N^{-1+\tilde b_1-
\tilde b_2/2}$, but as explained at the end of the introduction,
these two factors are equal.}                                  

\subsec{The Partition Function As A Vector In Hilbert Space}

Now we will make a change in viewpoint, which is suggested by experience
with rational conformal field theory in two dimensions.
A rational conformal field theory on a Riemann surface $\Sigma$ of positive
genus generally has, if one considers the chiral degrees of freedom only,
not a single partition function, but a collection of partition functions.  
It is useful \ref\shenkerfriedan{D. Friedan and S. Shenker, ``The
Analytic Geometry Of Two-Dimensional Conformal Field Theory,''
Nucl. Phys. {\bf B281} (1987) 509.} to group these
together as a vector in a Hilbert space, which one 
can think about using quantum mechanical intuition
\ref\verlinde{E. Verlinde, ``Fusion Rules And Modular
Transformations In $2-D$ Conformal Field Theory,'' Nucl. Phys.
{\bf B300} (1988) 360.}, and which one can
ultimately understand using topological
field theory in one dimension higher \ref\wittenjones{E. Witten,
``Quantum Field Theory And The Jones Polynomial,''
Commun. Math. Phys. {\bf 121} (1989) 351.}.

\def\H{{\cal H}}
So we introduce a Hilbert space ${\cal H}$ with one orthonormal basis vector
$\Psi_w$ for every $w\in H^2(M,\Z_N)$.  We regard the $Z_w(\tau)$ as
components of a vector $\Psi(\tau)\in \H$, with $\Psi(\tau)=\sum_wZ_w(\tau)
\Psi_w$.  Thus
\eqn\ixoc{Z_w(\tau)=\bigl(\Psi_w,\Psi(\tau)\bigr).}

Since the gauge theory ``partition function'' is thus not an ordinary
function but takes values in the Hilbert space ${\cal H}$, the
AdS/CFT correspondence can only make sense if it is similarly true
that the Type IIB partition function on a manifold such as $X\times \S^5$
with conformal boundary $M$ is not an ordinary function but takes values
in ${\cal H}$.  

\def\O{{\cal O}}
How can this be?  
At this point, we must recall the general structure of the AdS/CFT
correspondence.

A very general class of observables in quantum field
theory is the following. Let  the $\O_i$ be a basis for the space of local
gauge-invariant 
operators, and let $J_i$ be $c$-number sources that 
couple to them.  Thus the generating functional of correlation functions
is
\eqn\pixoc{Z(\tau;J_i)=\left\langle\exp\left(\sum_i\int_MJ_i\O_i\right)\right
\rangle,}
where $\langle~~\rangle$ denotes the (unnormalized) expectation value.
The usual claim in the AdS/CFT correspondence
is that to compute via string theory on $X\times \S^5$
this generating functional in the conformal field theory
on $M$, one must fix the values of
fields on $X$ -- near its boundary -- to an asymptotic behavior
determined by the $J_i$.
Hence for fixed $J_i$, one usually claims that the boundary behavior of the
fields on $X\times \S^5$ are fixed; the partition function is then
a ``number,'' that is a function of the $J_i$.
To resolve our present conundrum, 
we must show that if $H^2(M,\Z_N)\not= 0$, then even after
specifying the values of all sources $J_i$ for all local operators $\O_i$
on $M$, the boundary values of the fields on $X\times \S^5$ are not completely
determined; and the dependence on the extra data, whatever it is, must
be such that the partition function for given $J_i$ is not a number but
a vector in $\H$. 

The reason that this is so is that
fixing the behavior of all of the local gauge-invariant observables
near the boundary of $X\times \S^5$ does not completely specify the
gauge-invariant data near the boundary.  There is global gauge-invariant
information that cannot be measured locally, because the Type IIB theory
has the two two-form fields $B_{RR}$ and $B_{NS}$.  One can always
add to any given $B$-field a flat $B$-field, without affecting any local
gauge-invariant information.  Flat $B$-fields on a spacetime such as
$Y=X\times \S^5$
are classified modulo gauge transformations
by $H^2(Y,U(1))$.  Near the boundary, this reduces to
$H^2(M,U(1))$. For $H^2(M,\Z_N)$ to be nonzero (with $M$ of dimension four),
$H^2(M,U(1))$ must likewise be nonzero.  Hence, our puzzle 
arises only when there are
flat $B$-fields near $M$, in which case 
the partition function of the string theory on $X\times \S^5$ depends
on additional data and not only on the sources $J_i$ of gauge-invariant
fields.

Roughly speaking, the additional data can be measured by ``$\theta$ angles''
\eqn\unx{\eqalign{\alpha_{NS} & =\int_S B_{NS} \cr
                  \alpha_{RR} & =\int_S B_{RR},\cr}}
with $S$ a homologically nontrivial two-dimensional surface in $M$.
One might think that the partition function of the string theory on
$X\times \S^5$ would be a function of these theta angles, as well
as the sources $J_i$.  That would be an interesting result, but not
quite what we need.  We want instead to see the finite group
$H^2(M,\Z_N)$.  The reason that the string theory partition function
should not be regarded simply as a function of $\alpha_{NS}$ and $\alpha_{RR}$
is that these are, in a sense, canonically conjugate variables.                    

The low energy effective action for $B_{RR}$, $B_{NS}$, after reduction
on $\S^5$, looks something like
\eqn\refro{L=-{iN\over 2\pi}\int_X B_{RR}\wedge dB_{NS}+
{1\over 2\gamma}\int_X \vert dB_{RR}\vert^2
+{1\over 2\gamma'}\int_X|dB_{RR}|^2 +\dots.}
Here the first term is the Chern-Simons term, whose significance for
understanding the role of the center of the gauge group in the AdS/CFT
correspondence was already mentioned in the introduction.  The second
and third terms (with constants $\gamma,\gamma'$ that depend on $\tau$)
are the conventional kinetic terms for two-form fields.  The ``$\dots$'' 
are additional gauge-invariant terms of higher dimension.

The first term in \refro\ -- the Chern-Simons term --
is the important one for our present purposes,  for several reasons.
The obvious reason is that this is the unique term with only one derivative
and hence dominates at long distances (recall that the conformal boundary
of $X$ is ``infinitely far away,'' so the behavior near the boundary
is a question of long distance physics). Perhaps even more fundamentally,
the second and third terms in \refro\ and all higher
terms are integrals of gauge-invariant local densities and hence insensitive
to the $\alpha$'s, which contribute only to the first term.
The Chern-Simons term is gauge-invariant but is not the integral of
a gauge-invariant local density; it can and in general does
change if one shifts the $B$-fields by a flat $B$-field (a fact that
was crucial in \ahaw\ to enable the expected thermal symmetry of $SU(N)$
gauge theory at nonzero temperature to emerge from the AdS/CFT correspondence).

So to address the question of $\alpha$-dependence of the partition function,
we can use near the boundary of $X\times \S^5$ the simplified
Chern-Simons action
\eqn\kulk{L_{CS}=NI_{CS}=-{iN\over 2\pi}\int_X B_{RR}\wedge dB_{NS}.}
If we write $H=dB$ for the field strength of a $B$-field, then
the equations of motion derived from the Chern-Simons action are 
$H_{RR}=H_{NS}=0$,
so this action governs only the $\alpha$-dependence (not the 
modes of nonzero $H$, which are massive in spacetime and
have been integrated out to reduce to \kulk; their
behavior near the boundary of $X\times \S^5$ is determined in the usual
way by the sources $J_i$ of gauge-invariant local operators on the boundary).

The Chern-Simons action $L_{CS}$ does not depend on a metric on $X$,
so the theory governing the $\alpha$-dependence is a topological field theory,
of a familiar kind \aschwarz.
From the first-order form of $L_{CS}$, we see that $B_{RR}$ and $B_{NS}$ are
canonically conjugate variables in this topological field theory.
  After imposing the equations of motion
and dividing by gauge transformations, this means that $\alpha_{RR}$ and
$\alpha_{NS}$ are canonically conjugate.  Hence, we should not attempt
to compute the partition function as a function of both $\alpha_{RR}$
and $\alpha_{NS}$.

What we should do instead follows from general concepts of quantum mechanics.
Near $M$, $X$ looks like $M\times \R$, with $\R$ the ``time'' direction.
By quantizing the Chern-Simons theory on $M\times \R$, we obtain
a ``quantum Hilbert space'' $\H'$ associated with $M$.  We should
interpret the partition function of the theory on $X$ as determining,
not a number, but a vector in $\H'$.  Because $\H'$ is obtained
by quantizing the metric-independent Chern-Simons Lagrangian, it depends only
on the topology of $M$, and not on a metric.

Moreover, the topological field theory with action $L_{CS}$ has
an $SL(2,\Z)$ symmetry, acting on the pair
\eqn\durko{\left(\matrix{B_{RR} \cr B_{NS}\cr}\right)}
in the standard fashion. (A subtlety concerning this assertion
will be explained in section 3.4.)
The Hilbert space $\H'$ hence has a natural
action of $SL(2,\Z)$.

This is almost what we need: to agree with the expectations in the boundary
gauge theory, the partition function of the string theory on $X\times \S^5$
should be (once the gauge-invariant boundary data are specified) not a number
but a vector in a Hilbert space $\H$.  
$\H$ is determined purely topologically by $M$
(by definition, it has an orthonormal basis consisting of vectors
 $\Psi_w$ for each $w\in H^2(M,\Z_N)$),
and has an action of $SL(2,\Z)$ that we described in section 2.1.

So all will be well if $\H=\H'$.  
Actually, we must describe somewhat more precisely in what sense these
two spaces should coincide.  The description of the space $\H$ via
its basis $\Psi_w,\,\,w\in H^2(M,\Z_N)$ is not invariant under $S$-duality
(which as seen in \fourtran\ does not preserve this basis).  To make this
description, we need to pick a notion of what we mean by ``magnetic flux,''
as opposed to ``electric flux.''  Such a choice breaks $SL(2,\Z)$.
$SL(2,\Z)$ is broken in the same way if we introduce a two-dimensional
lattice $\Lambda=\Z^2$ on which $SL(2,\Z)$ acts in the standard
way, and pick a ``polarization'' of
$\Lambda$.  
One can think of a polarization as a choice of one direction in the lattice
$\Lambda$.
Alternatively, one can think of the pair $(\alpha_{RR},\alpha_{NS})$
as taking values in $\R^2/\Lambda$; a polarization then is
just a choice of what integral
linear combination of $\alpha_{RR}$ and $\alpha_{NS}$
is the ``momentum'' variable.  So we must show that for every choice of
polarization, $\H'$ acquires a basis in one-to-one correspondence
with the elements of $H^2(M,\Z_N)$, and that $SL(2,\Z)$ acts on $\H'$
as it does on $\H$.

These assertions  will be demonstrated
in section 3.

\newsec{Quantization Of The Topological Field Theory}

\subsec{Preliminaries}

To quantize the Chern-Simons theory on $M\times \R$, we first
work out the gauge-invariant classical phase space.   We work in the gauge
(analogous to $A_0=0$ gauge for gauge theory) in which $i\,0$ components
of $B_{RR}$ and $B_{NS}$ ($i$ and $0$ label directions tangent to $M$ and 
$\R$, respectively) vanish.  In quantization, we restrict to time zero;
together with the gauge choice, this means that $B_{RR}$ and $B_{NS}$ are 
$B$-fields on $M$.  
The canonical commutation relations, if written out in detail in
components, read
\eqn\xkon{\left[B_{RR\,ij}(x),B_{NS\,kl}(y)\right]=-{2\pi i\over N}
\epsilon_{ijkl}\delta^4(x,y),}
with $[B_{RR}(x),B_{RR}(y)]=[B_{NS}(x),B_{NS}(y)]=0$.  
Here, $x$ and $y$ are points in $M$,
and $i,j,k,l=1\dots 4$ are indices tangent to $M$.


The ``Gauss's law'' constraint ($\delta L/\delta B_{i0}=0$,
where $B$ is $B_{RR}$ or $B_{NS}$) gives $H_{RR}=H_{NS}=0$.
So, modulo gauge transformations, the phase space consists of
pairs of flat $B$-fields. 

Flat $B$-fields are classified by $H^2(M,U(1))$.
$B$-fields in general are classified topologically by the characteristic
class $[H]\in H^3(M,\Z)$.  However, for a flat $B$-field, this characteristic
class vanishes as a differential form, and hence represents a
{\it torsion} element of $H^3(M,\Z)$.  Let $H^3(M,\Z)_{tors}$ be the
torsion subgroup of $H^3(M,\Z)$.  To any given flat $B$-field we can,
without changing the topological type, add a flat and topologically
trivial $B$-field, via a transformation $B\to B+\beta$, where $\beta$
is a globally-defined, closed two-form.  We should consider $\beta$ trivial
if its periods are multiples of $2\pi$ (since the periods of $B$ can be
shifted by multiples of $2\pi$ by a gauge transformation), 
so we regard $\beta$ as an element
of $H^2(M,\R)/2\pi H^2(M,\Z)$.  The space $H^2(M,U(1))$ of flat $B$-fields
thus fits into an exact sequence
\eqn\kixo{0\to H^2(M,\R)/2\pi H^2(M,\Z)\to H^2(M,U(1))\to H^3(M,\Z)_{tors}
\to 0.}
This says that a flat $B$-field has a characteristic class in 
$H^3(M,\Z)_{tors}$, and two flat $B$-fields with the same characteristic
class differ by an element of $H^2(M,\R)/2\pi H^2(M,\Z)$, which classifies
topologically trivial flat $B$-fields.

Topologically trivial flat $B$-fields can be measured by their
periods, which are the ``world-sheet theta angles,''
\eqn\icco{\eqalign{\alpha_{NS} & =\int_S B_{NS} \cr
                  \alpha_{RR} & =\int_S B_{RR},\cr}}
where $S$ is a closed two-surface in $M$.

There are thus two parts of the quantization: to quantize the $\alpha$'s,
and to take account of the finite group $H^3(M,\Z)_{tors}$.  
These turn out to present
quite different problems and a direct treatment of the quantization
involves many subtle details.  On the other hand, everything we need
to know can be deduced from the symmetries of the problem.
In section 3.2, we will consider this analysis using the symmetries.
Then -- for the benefit of curious readers -- we enter into
a direct analysis of the quantization.

\subsec{Symmetries}

The most obvious symmetry of the problem is that we can add to $B_{RR}$
or $B_{NS}$ any $B$-field $B'$ such that $NB'$ is pure gauge.  For
instance, under 
\eqn\ginos{B_{RR}\to B_{RR}+B',} the Chern-Simons action 
$L_{CS}(B_{RR},B_{NS})
=NI_{CS}(B_{RR},B_{NS})$
changes by $L_{CS}\to L_{CS}+NI_{CS}(B', B_{NS})=L_{CS}+I_{CS}(NB',B_{NS})$.
But $I_{CS}(NB',B_{NS})=0$ as $NB'$ is pure gauge.

For $NB'$ to be pure gauge means necessarily that $B'$ is flat.  Thus,
$B'$ determines an element of $H^2(M,U(1))$ that is ``of order $N$,''
or in other words is annihilated by multiplication by $N$.  

We can be more explicit about the quantum field operators that
generate these symmetries.  (They are analogs of operators
considered many years ago in two-dimensional rational conformal
field theory \verlinde.)  We construct them using the two-form counterparts
of the familiar Wilson and 't Hooft loop operators of gauge theories.

First we consider the counterparts of Wilson loops.
Let $S$ and $T$ be closed
two-surfaces in $M$, and let
\eqn\jsno{\eqalign{\Phi_{RR}(S)=&\exp\left(i\int_S B_{RR}\right),\cr
                   \Phi_{NS}(T)=&\exp\left(i\int_T B_{NS}\right).\cr}}
Since these are gauge-invariant operators, they act on the classical
phase space and map flat $B$-fields to flat $B$-fields.
From the canonical commutation relations, we can deduce that
\eqn\bsno{\Phi_{RR}(S)\Phi_{NS}(T)=\Phi_{NS}(T)\Phi_{RR}(S)\exp\left(
\left({2\pi i\over N}\right) S\cdot T\right),}
where $S\cdot T$ is the intersection number of the oriented surfaces
$S$ and $T$.  

In particular, while $\Phi_{RR}(S)$ commutes with $B_{RR}$, it shifts
$B_{NS}$ by a flat $B$-field that is essentially determined by \bsno\                   
to be ``Poincar\'e dual'' to $S$.  
In other words, we interpret the relation $\Phi_{RR}(S)\Phi_{NS}(T)
\Phi_{RR}(S)^{-1}=\Phi_{NS}(T)\exp(2\pi i S\cdot T/N)$ to mean
that conjugation by $\Phi_{RR}(S)$ shifts $B_{NS}$ by a flat $B$-field
with delta-function support on $S$ (this assertion is in any case
clear from the canonical commutation relations), thus multiplying
$\Phi_{NS}(T)$ by a phase.  The $1/N$ in the exponent in \bsno\ means
that $\Phi_{NS}$ is shifted by a flat $B$-field of order $N$.
Conversely, conjugation by $\Phi_{NS}(T)$ shifts
$B_{RR}$ by such a field.  Thus, these operators
generate the symmetries that we exhibited directly in \ginos.
The $c$-number factor in the commutation relation \bsno\ shows
that these symmetries
do not commute with one another; there is a central extension by the
$N^{th}$ roots of unity.

The operator $\Phi_{RR}(S)$, in acting on gauge-invariant states, is trivial
if $S$ consists of $N$ copies of another closed surface $S'$.  That is
because in such a case, $\Phi_{RR}(S)=\Phi_{RR}(S')^N$;
but $\Phi_{RR}(S')$ shifts $B_{NS}$ by a $B$-field of order $N$, and hence
$\Phi_{RR}(S')^N$ shifts it by a pure gauge.
Likewise, $\Phi_{NS}(T)$ is trivial if $T=NT'$ for some $T'$.

$\Phi_{RR}(S)$ (or $\Phi_{NS}(T)$) is also trivial if $S$ (or $T$) is
the boundary of a three-manifold $U\subset M$.  For
then
\eqn\osno{\Phi_{RR}(S)=\exp\left(i\int_U H_{RR}\right) ,}
and this is trivial as an operator on gauge-invariant states
because of the Gauss law constraint $H=0$.

So the operators $\Phi_{RR}(S)$ depend on $S$ modulo boundaries, and
are trivial if $S$ is of the form $NS'$.  So far we have assumed that
$S$ has no boundary, but we will next see that we can define an operator
$\Phi_{RR}(S)$ with similar properties if $S$ has a boundary,
provided this boundary is of the form $NC$ for
some circle $C\subset M$.  This will make possible the following
simple description of the symmetry group.
The group $H_2(M,\Z_N)$ classifies two-surfaces
$S\subset M$ with boundaries of the form $NC$, modulo those of the form
$NS'$ and modulo boundaries.  So once we show that $S$ can have a boundary
of the claimed kind, we will have shown that the operators $\Phi_{RR}(S)$
(and similarly the operators $\Phi_{NS}(T)$) are classified 
by $H_2(M,\Z_N)$.  Equivalently, by Poincar\'e duality, they are classified
by $H^2(M,\Z_N)$.

\bigskip\noindent{\it 't Hooft Loops}

To see how $S$ can have a boundary, we begin with a rather
different-sounding question.
Are there also in this theory symmetry operators that are analogous
to the 't Hooft loops of gauge theory?  This would mean the following.  We 
consider a circle $C\subset  M$.  Deleting $C$ from $M$, we make
a ``singular gauge transformation'' on $B_{NS}$ (or $B_{RR}$) such
that on a small three-sphere  that links once around $C$,
 the integral of $H_{NS}$ equals $2\pi$.  This means
that we put on $C$ a ``magnetic source'' for $B_{NS}$.

To find such operators, we can go back to Type IIB superstring theory
on $X\times \S^5$.  In this theory, the NS fivebrane is a magnetic
source for $B_{NS}$, so the object sought in the last paragraph would
be a fivebrane wrapped on $C\times \S^5$.  However, as explained in
\bary, such a fivebrane must be the boundary of $N$ $D$-strings.
In other words, in addition to the magnetic loop, a $D$-string must
be present with a world-volume $S$ whose boundary consists of $N$ copies
of $C$.  

The operator $\Phi_{RR}(S)$ describes the coupling of $B_{RR}$ to the
$D$-brane worldvolume $S$.  So we have learned that $S$ can have a boundary
of the form $NC$, provided that there is at the same time a magnetic
source for $B_{NS}$ on $S$.

The recourse to string theory to show the existence of mixed
Wilson/'t Hooft operators of this kind is a convenient shortcut, but of course
it would be desireable to demonstrate their existence directly
in the five-dimensional topological field theory.  In fact, this
has essentially already been done in section 5 of \googuri,
in the course of explaining a low energy
approach to the existence of the baryon
vertex in Type IIB on $AdS_5\times \S^5$.

\bigskip\noindent{\it The Heisenberg Group And The Hilbert Space}

The operators  $\Phi_{RR}(S)$ (and likewise $ \Phi_{NS}(T)$) are
thus classified by $H_2(M,\Z_N)$ or equivalently by $K=H^2(M,\Z_N)$. 
We write these operators now as $\Phi_{RR}(w)$, $\Phi_{NS}(w)$, with
$w\in K$.  The commutation relation of these operators is still given
by \bsno.  We can assume that all intersections of $S$ and $T$ occur
away from the boundaries, so there is no real modification in the derivation
of \bsno\ from the canonical commutation relations.  Since $S$ and $T$
may have boundaries of the form $NC$, the intersection number $S\cdot T$
is only well-defined modulo $N$, but that is good enough to make
sense of the phase factor in the commutation relation.

Because of the central factor in 
the commutation relation \bsno, the group generated by these operators
is not just $K\times K=H^2(M,\Z_N)\times H^2(M,\Z_N)$, but is  a central extension
of $K\times K$ by $\Z_N$, the group of $N^{th}$
roots of unity.  We call this central extension $W$:
\eqn\oxnon{0\to \Z_N\to W\to K\times K\to 0.}
This central extension is nondegenerate (a statement which, informally,
means that the first or second factors of $K$ in the product $K\times K$
give maximal commutative subgroups of $W$) and is a group of a type
known as a finite Heisenberg group.  

A ``polarization'' of such a finite
Heisenberg group is a choice of a maximal commutative subgroup of $K\times K$,
or more exactly a maximal subgroup that remains commutative when lifted
to $W$.  For example, the first or second factor of $K\times K$,
or any ``diagonal'' subgroup obtained from then by an $SL(2,\Z)$
transformation, is a polarization.  There are also, in general, other
polarizations.  Picking a polarization of a finite Heisenberg group
is a discrete version of picking a representation of the canonical
commutation relations $[p,x]=-i\hbar$ for bosons.

One of the main theorems about such finite Heisenberg groups is that,
up to isomorphism, such a group has a unique irreducible representation $R$.
This is a discrete version of the fact that the canonical commutation
relations  for bosons have a unique irreducible representation, up to
isomorphism.  The representation can be constructed as follows.
Since the $\Phi_{RR}(w)$ commute, one can pick a basis of $R$ consisting
of their joint eigenstates.  Using the nondegeneracy, one can show
that there is a vector $|\Omega\rangle\in R$ that is invariant under
all of the $\Phi_{RR}(w)$.  From irreducibility of $R$, one can show
that $|\Omega\rangle$ is unique (up to multiplication by a scalar) and
that $R$ has a basis consisting of the states
\eqn\inoop{\Psi_w=\Phi_{NS}(w)|\Omega\rangle,~~ {\rm for}~w\in H^2(M,\Z_N).}
The action of the group in this basis follows immediately from the
commutation relations.

If the quantum Hilbert space $\H'$ of the Chern-Simons theory
is an {\it irreducible} representation of the finite symmetry group
$W$, then it is easy to obtain all the properties promised in section 2.2.
For example, we must show that for each polarization 
of the lattice $\Lambda=\Z^2$ on which $SL(2,\Z)$ acts,
$\H'$ has a basis $\Psi_w,\,w\in H^2(M,\Z_N)$. In this context, a polarization
is a choice of what we mean by $\Phi_{RR}$ and what we mean by $\Phi_{NS}$.
The desired basis is constructed
in \inoop\ for one choice of polarization, and other polarizations are
obtained by $SL(2,\Z)$ transformations.
There is indeed a natural
action of $SL(2,\Z)$ on $\H'$, since the commutation relation \bsno\ by
which $W$ is defined is $SL(2,\Z)$-invariant.  It is not hard to see
that the $SL(2,\Z)$ action agrees with what was described in section 2.1.
The main point is that the $SL(2,\Z)$ transformation 
\eqn\sonx{\left(\matrix{ 0 & 1 \cr -1 & 0 \cr}\right),}
which acts by $\tau\to -1/\tau$,
corresponds to  a change in polarization that maps $\Phi_{RR}$
to $\Phi_{NS}^{-1}$ and $\Phi_{NS}$ to $\Phi_{RR}$.  It maps the
state $|\Omega\rangle$ invariant under the $\Phi_{RR}$'s to the
state 
\eqn\cxo{|\tilde \Omega\rangle={1\over N^{\tilde b_2/2}}\sum_{w\in K}
      \Phi_{NS}(w)|\Omega\rangle,}
which is invariant under the $\Phi_{NS}$'s.  So it maps $\Psi_v=
\Phi_{NS}(v)|\Omega\rangle$ to 
\eqn\kxonxz{\Phi_{RR}(v)|\tilde \Omega\rangle={1\over N^{\tilde b_2/2}}
\Phi_{RR}(v)\sum_{w\in K}\Phi_{NS}(w)|\Omega\rangle={1\over N^{\tilde b_2/2}}
\sum_{w\in K}\exp\left(2\pi i v\cdot w/N\right)\Phi_{NS}(w)|\Omega\rangle.}
This is the discrete Fourier transform familiar from section 2.1.
The transformation under $\tau\to\tau+1$ can be understood similarly.
 
Except for the fact that we have not shown that the finite group $W$
acts irreducibly in the quantum theory, this really establishes what we wanted.
However, we will go on in the rest of this section, for the benefit
of curious or intrepid readers, to show what is involved in quantizing
the theory directly.  In the process, it will become clear
that $W$ acts irreducibly on $\H'$.

In the analysis, it will be useful to know the following more general
description of the representation $R$.  First of all, if $F$ is any
polarization of $K\times K$, then $R$ has a basis that is in
one-to-one correspondence with the cosets of $(K\times K)/F$.  This is proved
by first showing that $R$ has an $F$-invariant vector $|\Omega\rangle$,
and then setting $\Upsilon_\lambda=\Phi_\lambda |\Omega\rangle$, where
$\lambda$ runs over a set of representatives 
of the cosets of $K/F$, and for each $\lambda$, $\Phi_\lambda$
is the corresponding operator on $R$.
The $\Upsilon_\lambda$ are the desired basis vectors.

\subsec{The Torsion-Free Case; Elementary Account}

Direct quantization of this system is surprisingly subtle, given
that it is an abelian free field theory.  The reader may in fact
wish to jump
directly to section 4.  

We begin by considering
the case that there is no torsion in $H^2(M,\Z)$.   
If $H^2(M,\Z)$ is torsion-free, then this group is a lattice $\Gamma$.
The intersection pairing on $H^2(M,\Z)$ gives an integer-valued
inner product on $\Gamma$; we write the product of $x,y\in \Gamma$
as $x\cdot y$.  This inner product is unimodular by Poincar\'e duality
and is even because (as we noted at the outset) $M$ is spin.

The exact sequence of abelian groups $0\to \Z\underarrow{N}\Z\to \Z_N\to 0$
leads to a long exact sequence of cohomology groups which reads in part
\eqn\pxi{\dots H^2(M,\Z)\underarrow{N} H^2(M,\Z)\to H^2(M,\Z_N)\to 
H^3(M,\Z)\underarrow{N} H^3(M,\Z)\dots.}
If there is no torsion in the cohomology of $M$, then the map
$H^3(M,\Z)\underarrow{N} H^3(M,\Z)$ is injective.  Hence exactness of \pxi\
says in the torsion-free case that
\eqn\kuji{H^2(M,\Z_N)=H^2(M,\Z)/NH^2(M,\Z)=\Gamma/N\Gamma.}
In other words, in the absence of torsion, a $\Z_N$ class is the
mod $N$ reduction of an integral class.

In the torsion-free case, the world-sheet theta angles $\alpha_{NS}$ 
and $\alpha_{RR}$ are a complete set of gauge-invariant functions
on the classical phase space.  They are canonically conjugate variables,
a fact that we have already  used in the analysis of the symmetries
in section 3.2.
A precise form of this statement
is that if we pick  for the lattice $\Gamma$ a basis in which the inner
product is $g_{ij}$, $1\leq i,j\leq b_2$,  and write $\alpha^i_{NS}$,
$\alpha^i_{RR}$ for the corresponding components of $\alpha_{NS}$ and
$\alpha_{RR}$, then the Poisson brackets are 
\eqn\tuggo{\{\alpha_{RR}^i,\alpha_{NS}^j\}={2\pi\over N}g^{ij}.}

According to the description at the end of section 2.2, we are to 
pick a polarization, quantize, and determine the resulting Hilbert
space ${\cal H}'$.  We pick a polarization\foot{
For some background on quantization of Chern-Simons
theories, see
\nref\canqu{S. Elitzur, G. Moor, A. Schwimmer, and N. Seiberg,
``Remarks On The Canonical Quantization Of Chern-Simons-Witten
Theory,'' Nucl. Phys. {\bf B326} (1989) 108.}
\nref\axetal{S. Axelrod, S. Della Pietra, and E. Witten, 
``Geometric Quantization Of Chern-Simons Gauge Theory,''
 J. Diff. Geom. {\bf 33} (1991) 787.}
\refs{\canqu,\axetal}.}
 by regarding $\alpha_{NS}$
as the ``position'' variable and
\eqn\imon{\beta_{RR}={N\over 2\pi}\alpha_{RR}}
as the conjugate momentum.  Note that as the periods of $\alpha_{RR}$
are defined mod $2\pi$, those of $\beta_{RR}$ are defined mod $N$,
so classically $\beta_{RR}$ is an element of $H^2(M,\R)/NH^2(M,\Z)=V/N\Gamma$,
where $V=H^2(M,\R)$ and $\Gamma=H^2(M,\Z)$ is regarded as a lattice in $V$.  

Quantum mechanically there is a very severe additional constraint
on $\beta_{RR}$.
Since the ``position'' variable $\alpha_{NS}$ is a periodic variable
that takes values in $H^2(M,\Z)/2\pi \Gamma$, the conjugate
momentum $\beta_{RR}$ takes values in the  lattice $\Gamma^*$ dual to $\Gamma$.
But Poincar\'e duality tells us that $\Gamma^*=\Gamma$, so $\beta_{RR}$
takes values in $\Gamma$.  Since $\beta_{RR}$ is in addition defined
modulo $N\Gamma$, it takes values in $\Gamma/N\Gamma$.  According to
\kuji, this is the same as $H^2(M,\Z_N)$.

So we get the desired result.  There is one quantum state -- one
momentum state $\Psi_w$ -- for every momentum vector $w\in \Gamma/N\Gamma
=H^2(M,\Z_N)$.

Moreover, the 
$SL(2,\Z)$ transformation $\tau\to -1/\tau$ exchanges the position
$\alpha_{NS}$ with the momentum $\alpha_{RR}$.  Hence it acts
as a Fourier transform, in agreement with the gauge theory result
\fourtran.

\subsec{More Rigorous Approach}

\def\M{{\cal M}}
\def\L{{\cal L}}
The  discussion in section 3.3 was
 actually somewhat informal, and for the interested
reader we will now give some hints for a more precise treatment.

Let $T$ be the torus $T=V/\Gamma$ with $V=H^2(M,\R)$ and $\Gamma=H^2(M,\Z)$.
The phase space of the system is $T'=T\times T$.  The first step
in quantizing the theory is to find the appropriate line bundle 
$\M$
over the phase space.  $\M$ should be endowed with a connection
whose  curvature equals  the symplectic two-form of the theory.
Once the right line bundle is found, quantization is carried out
by picking a complex structure and taking holomorphic sections of
$\M$, or by using a real polarization and making a more precise version
of the informal discussion of section 3.3, or by finding and using
a finite Heisenberg group as in section 3.2.  But any approach
to quantization involves finding the line bundle at the outset.

Since $N$ appears in the symplectic form as a multiplicative factor,
$\M$ will be of the form $\L^N$, where $\L$ is the line bundle
that we would get at $N=1$.  (The relation $\M=\L^N$ will be obvious
from the Chern-Simons construction of the line bundle that we give
presently.)  

\nref\mikk{J. Mickelsson, ``Kac-Moody Groups, Topology Of The
Dirac Determinant Bundle, and Fermionization,'' Commum. Math. Phys.
{\bf 110} (1987) 173.}
\nref\ramadas{T. R. Ramadas, I. M. Singer, and J. Weitsman,
``Some Comments on Chern-Simons Theory,'' Comm. Math. Phys.
{\bf 126} (1989) 409.}
\nref\axel{S. Axelrod, {\it Geometric Quantization Of Chern-Simons
Theory,} Ph.D. Thesis, Princeton University (1991).}
\nref\freed{D. Freed, ``Classical Chern-Simons Theory: Part I,''
Adv. Math. {\bf 113} (1995) 237.}
The construction of the line bundle is most naturally made
directly from the Chern-Simons action \refs{\mikk - \freed}.
Let $p$ be a point in the classical phase space $T'$ given
by a pair $(B_{RR},B_{NS})=(a,b)$. Let
$0$ be the origin in $T'$ (the point with $B_{RR}=B_{NS}=0$).
Then we describe the fiber $\L_p$ of $\L$ at $p$ as follows.
For any path $\gamma$ in $T'$ from $0$ to $p$, $\L_p$
has a basis    vector $\psi_\gamma$, of norm 1.  If
$\gamma$ and $\gamma'$ are two such paths, we declare
that 
\eqn\jumbo{\psi_{\gamma'}=e^{iL_{CS}}\psi_\gamma,}
where the Chern-Simons Lagrangian $L_{CS}$ is understood in the following
sense.  The path $\gamma$ from $0$ to $p$ determines a ``time''-dependent
pair of $B$-fields $(B_{RR}(t),B_{NS}(t))$,
where $(B_{RR}(0),B_{NS}(0))=(0,0)$ and $(B_{RR}(1),B_{NS}(1))
=(a,b)$.  We can fit these together to make a pair
$(B_{RR},B_{NS})$ over $M\times I$, where $I$ is a unit interval.
Likewise, $\gamma'$ determines a pair of $B$-fields over $M\times I$.
By gluing $\gamma'$ to $\gamma$, with opposite orientation for $\gamma'$,
we get a $B$-field pair over the closed five-manifold $M\times \S^1$.
$L_{CS}$ in \jumbo\ is the Chern-Simons action evaluated for this field.

It can be shown that the line bundle $\L$, whose
fiber at each $p\in T'$ was just described,
is endowed with a natural connection,
whose curvature is the symplectic form.   
The monodromy $M_C$ of this line bundle around any circle $C\subset
T'$ is as follows.  The circle $C\subset T'$ determines a $B$-field pair
over $M\times C$, and the monodromy is
\eqn\eqmon{Q(C)=\exp(-L_{CS}),}
with $L_{CS}$ the Chern-Simons action of this pair.

Having found the line bundle, we are ready to quantize.
Quantum states are suitable
sections of $\M=\L^N$, with the details depending on a choice of polarization.

In the present case, we can make the description of $\L$ much more
explicit.  In general, a line bundle with connection over any manifold $Z$
is determined up to isomorphism by giving its monodromies around
arbitrary loops in $Z$.  Once the curvature is specified, it suffices
to consider a set of loops generating the fundamental group of $Z$.
In the present instance, $Z$ is the torus $T'=T\times T=(V\times V)/(\Gamma
\times \Gamma)$.  The fundamental group of $T'$ is generated by
straight lines from the origin in $V\times V$ to lattice points
in $\Gamma\times \Gamma$.  Let $(x,y)$ be such a lattice point.
A straight line from the origin to this lattice point is given by
the family of $B$-field pairs
\eqn\rurrgy{\eqalign{B_{RR} & = 2\pi tx \cr
                     B_{NS} & = 2\pi ty ,\cr}}
with $0\leq t\leq 1$. At $t=1$,
$(B_{RR},B_{NS})=(2\pi x,2\pi y)$ are gauge-equivalent to 0. So the
family $(B_{RR}(t),B_{NS}(t))$ gives a closed
loop in the phase space.  The monodromy around this loop can be computed by
direct evaluation of $L_{CS}=-(i/2\pi)\int_{M\times \S^1}B_{RR}\wedge dB_{NS}$,
\foot{To put this ``direct evaluation'' on a rigorous basis is slightly
subtle as the $B$-fields on $M\times \S^1$ are topologically nontrivial.
The basic idea of a rigorous evaluation in this situation is contained
in Example 1.16 in \cheeger.}
and is
\eqn\memm{Q(x,y)=(-1)^{x\cdot y}.}
The relation
\eqn\emmo{Q(x+x',y+y')=Q(x,y)Q(x',y')(-1)^{x\cdot y'+y\cdot x'}}
follows from this, and signals, as explained in connection with
eqn. (2.17) of \witto, that the line bundle ${\cal L}$ has the desired
curvature and first Chern class.

The formula for $Q(x,y)$ is clearly invariant under all of the lattice
symmetries of $\Gamma$, and thus under all diffeomorphisms of $M$.
Let us check that it is also invariant under $SL(2,\Z)$, acting
in the standard fashion
\eqn\opemm{\eqalign{ x\to ax+by\cr
                     y\to cx+dy\cr}}
with $a,b,c,d$ integers such that $ad-bc=1$.
A small computation shows that $Q(x,y)$ is invariant under this transformation
for all such  $a,b,c$, and $d$ if and only if $x^2$ and $y^2$ are both
even.  But the manifold $M$ is spin, and the intersection form
on $H^2(M,\Z)$, for $M$ a four-dimensional spin manifold, is
always even.  So $x^2$ and $y^2$ are even, and $Q$ has the expected symmetries.
Since $Q$ determines the line bundle ${\cal L}$, ${\cal L}$ has these
symmetries also.

At this point, the reader may wonder precisely why the spin condition
on $M$ was needed.   The line bundle ${\cal L}$ was determined directly
from the Chern-Simons action, which appears to be completely 
$SL(2,\Z)$-invariant without assuming that $M$ is spin;
so why did the spin structure enter?   The answer will make it clear
that up to the present point, we have indulged in a small sleight of hand.
There is actually a subtlety
in defining the Chern-Simons action for a pair of $B$-fields $(B_{RR},
B_{NS})$ on a five-manifold $X$, even if $X$ has no boundary.  
If $X$ is the boundary of a six-manifold
$Y$, and the $B$-fields extend over $Y$, then the action is readily
defined as
\eqn\buffo{I_{CS}=-{i\over 2\pi}\int_Y H_{RR}\wedge H_{NS}.}
This formula is completely $SL(2,\Z)$-invariant, proving that
$SL(2,\Z)$ is a symmetry if $Y$ exists.
If no such $Y$ exists, then as mentioned in the introduction,
one must define the action by
$\exp(-I_{CS})=B_{RR}*B_{NS}$, with $*$ the multiplication
in Cheeger-Simons cohomology.  From Theorem 1.11 of \cheeger,
one can deduce that $I_{CS}(B_{RR},B_{NS})=I_{CS}(B_{NS},-B_{RR})$ and
that $I_{CS}(B_{RR},B_{NS})=I_{CS}(B_{RR}+2B_{NS},B_{NS})$.  $SL(2,\Z)$ holds
if and only if one has the more precise property $I_{CS}(B_{RR},B_{NS})
=I_{CS}(B_{RR}+B_{NS},B_{NS})$, but this does not follow from the theorem
and is evidently not true in general if $X$ is not spin.  The $SL(2,\Z)$
invariance of the explicit formula \memm\ that determines the line
bundle shows that, at least in canonical quantization, there is no
such difficulty in the spin case.

\bigskip\noindent{\it Quantization}

Having thus defined the line bundle, we now wish to carry out quantization.
This may be done by picking a polarization and quantizing, but 
we prefer to make contact with the discussion of section 3.2 by
exhibiting the discrete Heisenberg group.

Since the phase space $T'$ is a torus, some obvious symmetries of the
phase space (as a symplectic manifold)
are translations of the torus, say $(B_{RR},B_{NS})\to (B_{RR},B_{NS})
+2\pi (a,b)$, with $a,b\in V/\Gamma$.  However, such translations
generally do not leave fixed the line bundle ${\cal L}$.  Under
the indicated translation, the monodromy $Q(x,y)$ computed above
transforms by
\eqn\pson{Q(x,y)\to \exp(2\pi i(a\cdot y-b\cdot x)) Q(x,y).}
Thus, the monodromies are invariant, for all $x,y$, only if $a,b\in \Gamma$,
in which case the translation by $2\pi (a,b)$ acts trivially on the torus
$T'$.

So if the quantum line bundle is $\L$, there are no nontrivial translation
symmetries.
We are more generally interested in the ``level $N$'' theory in which
the quantum line bundle is $\M=\L^N$.  In this case, the monodromies
are $Q^N$, and it is sufficient for $Q^N$ to be invariant.
For this, it is enough that $Na $ and $Nb$ should be lattice points.
So the translation by $2\pi (a,b)$ is a symmetry of the quantum theory
whenever $a$ and $b$ are both of order $N$.  The points of order $N$
are classified by ${1\over N}\Gamma/\Gamma=\Gamma/N\Gamma=H^2(M,\Z_N)$.

Let us try to define as precisely as possible the symmetry 
$T_{a,b}$ of translation by $2\pi (a,b)$.  Such a symmetry exists
because the pullback of $\M$ by the translation in question is isomorphic
to $\M$.  The operator $T_{a,b}$ is uniquely determined up to multiplication
by a complex scalar of modulus 1.  That scalar can be restricted
by requiring (since the translation by $2\pi (Na,Nb)$ is trivial)
that $T_{a,b}^N=1$.  This leaves an ambiguity consisting of multiplication
by $N^{th}$ roots of unity.  There is no natural way to fix that remaining
ambiguity, so we make any arbitrary choice.

One might think that the translation operators $T_{a,b}$ would commute.
That is not so, because in the presence of a magnetic field (the symplectic
form on $T'$ serves as the ``magnetic field'') translations do not commute.
Including in the standard way the effects of the magnetic field, the
actual relation is
\eqn\kxo{T_{a,b}T_{a',b'}=T_{a',b'}T_{a,b}\exp
\left(2\pi i(a\cdot b'-b\cdot a')/N\right).}
This is the Heisenberg group found in section 3.2. The quantum theory
can now be analyzed as in section 3.2, with the important difference that
we can now show that $W$ acts irreducibly.  For instance,
this follows by using a complex polarization and Riemann-Roch theorem
 to determine the dimension
of the quantum Hilbert space, which coincides with that of the irreducible
representation $R$ of $W$.  For that matter, the informal treatment in section
3.3 was precise enough, at least for large enough $N$, to count the
quantum states within a positive integer
factor, and thus to deduce that $W$ acts irreducibly.

\subsec{Restriction On $[H]$}

Now we want to begin to consider what happens if there is torsion in
the cohomology of $M$.

We first note that $B$-fields on $M$ are characterized topologically
by a characteristic class $[H]\in H^3(M,\Z)$.
The Gauss law constraint implies that $H=dB$ is zero as a differential
form, so $[H]$ must be torsion.  At first sight, it appears
that there is no other restriction, and that in quantizing
the Chern-Simons theory, $[H_{RR}]$ and $[H_{NS}]$ can be arbitrary torsion
elements of $H^3(M,\Z)$.

Actually, there is a very important further restriction.  This
is that if the Chern-Simons theory is taken at level $N$, then 
$[H_{RR}]$ and $[H_{NS}]$ must be $N$-torsion, that is
$N[H_{RR}]=N[H_{NS}]=0$.

To obtain this result, we will consider the partition function of the
theory on the five-manifold $X=M\times \S^1$.  In topological field
theory, this partition function equals the dimension of the space of physical
states on $M$, and so in particular every physical state contributes.

We consider classical configurations on $M\times \S^1$
in which, when restricted
to $M$ (that is to $M\times P$ for any $P\in \S^1$), $[H_{NS}]$
has some given value.  We want to show that the contribution to the 
path integral of such configurations is zero unless $N[H_{NS}]=0$.
The same argument, with $B_{RR}$ and $B_{NS}$ exchanged, shows
that the contribution is zero unless also $N[H_{RR}]=0$ when restricted
to $M$.

The vanishing will come upon adding to $B_{RR}$ a flat $B$-field  $B'$
that is topologically trivial if restricted to $M$, but nontrivial
on $M\times \S^1$. (We restrict $B'$ to be trivial on $M$
because we consider a path integral in which $[H_{RR}]$ and $[H_{NS}]$
are both specified on $M$; we want to show that summing over what
happens in the ``time'' direction gives the desired restriction on the
initial data.)
  Under $B_{RR}\to B_{RR}+B'$, the path integrand
$e^{-L_{CS}}$ of Chern-Simons theory transforms by
\eqn\pxkk{\exp({-L_{CS}})\to \exp({-L_{CS}})\exp\left(
{ i N\over 2\pi}\int_{M\times \S^1}B'\wedge dB_{NS}\right),}
where the exponential factor needs some clarification (which will be
given shortly) because  the $B$-fields in question
are topologically nontrivial, but is hopefully clear.  If the exponential
factor in \pxkk\ does not equal 1, then the path integral will vanish
after summing over $B'$.

We must therefore understand what is meant by the  factor
that we write symbolically as
\eqn\hjkk{\exp\left({i\over 2\pi}\int_XB'\wedge dB_{NS}\right),}
if $B'$ and $B$ are flat but topologically nontrivial.
This is determined in eqn. 1.14 of \cheeger; the intuitively
natural result can be explained as follows.
We can classify flat $B$-fields in five dimensions either by
$H^2(X,U(1))$ or by the characteristic class
$[H]\in H^3(X,\Z)$.  (The first classification describes the $B$-field
up to gauge transformation, and the second only describes its topological
type.)  
There is a natural, nondegenerate pairing
\eqn\kxo{H^2(X,U(1))\times H^3(X,\Z)\to H^5(X,U(1))=U(1)}
given by Poincar\'e and Pontryagin duality.  For $B'\in H^2(X,U(1))$
and $B_{NS}$ regarded as an element of 
$H^3(X,\Z)$, we write this pairing as $E(B',B_{NS})$.  
Cheeger and Simons show that in this situation $B'*B_{NS}= E(B',B_{NS})$,
so we must interpret the phase factor in \hjkk\ as $E(B',B_{NS})$.

Because of the factor of $N$ in the exponent \pxkk, the factor that must
be 1 for $B_{NS}$ to contribute to the path integral is actually
$E(B',B_{NS})^N=E(B',NB_{NS})$.  Nondegeneracy of the pairing \kxo\ means
that for this to equal 1 for all $B'$, we need $NB_{NS}=0$, up to gauge 
transformation.  In other words, we need $N[H_{NS}]=0$.
In our case with $X=M\times \S^1$, since we assume $B'=0$ when
restricted to $M$, the constraint is that $N[H_{NS}]=0$ when restricted
to $M$.  The is the restriction that we aimed to prove.

The restriction that we have found should be viewed as a quantum
extension of Gauss's law, to incorporate torsion.  Gauss's law enters
in path integrals as a constraint that states must obey, in order to
propagate in time.  Such propagation is precisely what we have been
analyzing.
The classical Gauss law says that $H_{RR}=H_{NS}=0$ as a differential
form; the quantum Gauss law says that in addition $N[H_{RR}]=N[H_{NS}]=0$.

\subsec{Quantization In The Presence Of Torsion}

We will now analyze the quantum Hilbert space $\H'$ of the Chern-Simons
theory, allowing for the possible existence of torsion in $H^2(M,\Z_N)$.
In fact, to clarify things, we will consider first the case that 
the cohomology of $M$ contains only torsion -- the opposite case
from what we considered in sections 3.3 and 3.4.

After imposing Gauss's law, the physical data consists of a pair
of flat $B$-fields $B_{RR}$, $B_{NS}$.  From what we have seen in
section 3.5, we should impose a discrete form of Gauss's law
stating that $N[H_{RR}]=N[H_{NS}]=0$.  Let $L_N$ be the subgroup
of $H^3(M,\Z)$ consisting of points of order $N$.  The two $B$-fields
determine  a pair of points in $L_N$.   If the cohomology of $M$
is pure torsion, there are no additional data to quantize.  Each pair
$(x,y)\in L_N\times L_N$ contributes one quantum state $\Psi_{x,y}$,
and the $\Psi_{x,y}$ give a basis of the physical Hilbert space $\H'$.

We want to compare this to the  description from section 2: for each
choice of polarization,
$\H'$ should have a basis $\Psi_w$, for $w\in K=H^2(M,\Z_N)$. 
In an equivalent version presented in section 3.2,
 $\H'$ should furnish an irreducible description of the
finite Heisenberg group $W$:
\eqn\sononso{0\to \Z_N\to W\to K\times K\to 0.}
 
  For this, we 
must understand the structure of $H^2(M,\Z_N)$ when there is torsion.
From the exact sequence of abelian groups
$0\to \Z\underarrow{N}\Z\to \Z_N\to 0$, with the first map being
multiplication by $N$ and the second reduction modulo $N$, we get
a long exact sequence
\eqn\overlong{\dots \to H^2(M,\Z)\underarrow{N} H^2(M,\Z)
\to H^2(M,\Z_N)\to H^3(M,\Z)\underarrow{N} H^3(M,\Z)\to\dots.}
This gives a short exact sequence
\eqn\rlong{0\to E_N\to H^2(M,\Z_N)\to L_N\to 0,}
with $E_N=H^2(M,\Z)/NH^2(M,\Z)$ the subgroup of $H^2(M,\Z_N)$ consisting
of classes that are the reduction of an integer class.

We recall that there is a nondegenerate pairing
$H^2(M,\Z_N)\times H^2(M,\Z_N)\to \Z_N$ given by Poincar\'e duality;
it was used in section 3.2 to construct the finite Heisenberg group.
If the cohomology of $M$ is purely torsion, then $E_N$ consists entirely
of classes that are reductions of torsion classes.  In this case,
the pairing on $H^2(M,\Z_N)\times H^2(M,\Z_N)$ is zero when restricted
to $E_N\times E_N$ (since in integer cohomology the cup product vanishes
for torsion classes), and the self-duality of $H^2(M,\Z_N)$ reduces
to a duality between $E_N$ and $L_N$.

The Heisenberg group $W$ in this situation has a maximal commutative
subgroup that comes from the subgroup $E_N\times E_N$ of $H^2(M,\Z_N)
\times H^2(M,\Z_N)$.  According to the remark at the end of section
3.2, the representation $R$ has therefore a basis in 1-1 correspondence
with the cosets in $(K\times K)/(E_N\times E_N)$.
That quotient is isomorphic to $L_N\times L_N$, so we get one
basis vector $\Psi_{x,y}$ for each pair $(x,y)\in L_N\times L_N$.
This is the description we found for the physical Hilbert space $\H'$,
so we confirm the expected isomorphism of $R$ with $\H'$.

In particular, we see again that $\H'$ is  an irreducible representation
of the discrete Heisenberg group.

\bigskip\noindent{\it The General Case}

The general case in which the cohomology of $M$ contains torsion but
is not purely torsion is a mixture of the cases that we have already
considered.

In this case, the cohomology classes of $x=[H_{RR}]$, $y=[H_{NS}]$ determine
elements $x,y\in L_N$ which are part of the data.  In addition,
one must quantize the $\alpha$'s.  We can pick a polarization
in which the quantum states are regarded as  
functions of $\alpha_{NS}$ as well as $x,y$.  

As we found in section 3.3,
quantization of the $\alpha$'s gives one state for every topologically
trivial flat $B_{NS}$-field of order $N$.  Including also the dependence on $y$
means that we drop the requirement that $B_{NS}$ should be topologically
trivial; we get one state for every flat $B_{NS}$ field of order $N$
whether it is topologically trivial or not.
We let $M_N$ be the group of flat $B$-fields of order $N$.  Including
also the choice of $x$ (which is a point in $L_N$), 
we see that the quantum Hilbert space $\H'$ has
a basis consisting of one vector for every element of $L_N\times M_N$.
This strongly suggests that there might be a polarization of the
discrete Heisenberg group determined by a subgroup 
$F$ of $K\times K$ such that $(K\times K)/F=L_N\times M_N$.
In that case, the irreducible representation $R$ of the discrete
Heisenberg group can be identified with $\H'$ as desired.

In fact, we can take $F=E_N\times T_N$, where $T_N$ is the subgroup
of $H^2(M,\Z_N)$ consisting of elements that are the reduction
mod $N$ of torsion classes in $H^2(M,\Z)$.  $F$ is a commutative subgroup
(even when lifted to $W$)
because the pairing of two elements in $H^2(M,\Z)$, one of which is torsion,
is zero.  Since $(K\times K)/F=(K\times K)/(E_N\times T_N)
=(K/E_N)\times (K/T_N)$,
the claim that  $(K\times K)/F=L_N\times M_N$ 
is equivalent to the existence of two exact sequences:
\eqn\twoex{\eqalign{0&\to E_N\to H^2(M,\Z_N)\to L_N\to 0,\cr
 0&\to T_N\to H^2(M,\Z_N)\to M_N\to 0.\cr}}  
 The first is in \rlong, and the second can be
derived from the sequence 
\eqn\jucxs{0\to \Z_N\to U(1)\underarrow{N} U(1)\to 0}
(the first map is the inclusion of the points of order $N$ in $U(1)$,
and the second is $a\to a^N$).
From this sequence we get a long exact cohomology sequence
\eqn\xonons{0\to H^1(M,U(1))/NH^1(M,U(1))  \to H^2(M,\Z_N)\to H^2(M,U(1))_N
\to 0.}
Here $H^2(M,U(1))_N$ is the kernel of $H^2(M,U(1))\underarrow{N}
H^2(M,U(1))$; this is the group of flat $B$-fields of order $N$, or
in other words is $M_N$.  Also,
$H^1(M,U(1))$ classifies flat $U(1)$ bundles.  Such a bundle
has a first Chern class which is a torsion element of $H^2(M,\Z)$;
and the map from  $H^1(M,U(1))/NH^1(M,U(1))$ to $H^2(M,\Z_N)$ is given
by mod $N$ reduction of the first Chern class.  Conversely,
every torsion element
of $H^2(M,Z)$ is the first Chern class of a flat line bundle.
So the image of $H^1(M,U(1))/NH^1(M,U(1))$ in $H^2(M,\Z_N)$ consists
precisely of the classes that are mod $N$ reductions of torsion classes,
or in other words this image is $T_N$.
This completes the explanation of the second exact sequence in \twoex.

\newsec{The $(0,2)$ Theory In Six Dimensions}

In this section, we will consider the analogous issues for the
$(0,2)$ conformal field theory in six-dimensions.
\foot{Some related issues involving a discrete flux in the $(0,2)$ model
have been discussed in \ref\origa{O. Ganor
and S. Sethi, ``New Perspectives On Yang-Mills Theories With
Sixteen Supersymmetries,'' JHEP {\bf 1} (1998) 7, hep-th/9712071.}.}
  We consider
only the $A_N$ theory, that is, the theory associated with $M$-theory
on $AdS_7\times \S^4$, with $N$ units of four-form flux on $\S^4$.
We will, in essence, be determining the analog for the $A_N$ case of
the (0,2) theory of the space of conformal blocks in two-dimensional
rational conformal field theory.  A difference from the problem considered
in sections 2 and 3 is that there is no obvious candidate from
classical geometry for what the answer should be.

To study the $(0,2)$ theory on a six-dimensional spin-manifold $M$,
we consider $M$-theory on spacetimes that look near infinity like
$X\times \S^4$, where $X$ has $M$ for conformal boundary.
By analogy with gauge theory in four-dimensions, we suspect that for
general $M$, the partition function will not be a number but a vector
in some finite-dimensional Hilbert space associated with $M$.

A mechanism analogous to what we studied in sections 2 and 3 immediately
presents itself.  The long wavelength theory on $X$ has a three-form
field $C$, whose effective action contains the Chern-Simons term
\eqn\pxol{L_{CS}=-i{1\over 2} {N\over 2\pi}\int_X C\wedge dC.}
If $N$ is odd, then because of the $1/2$ in \pxol, to make sense of
this theory requires a spin structure on $X$.
(This is explained in \witto, where the case $N=1$ was considered and the 
factor of $1/2$ in \pxol, which arises from a similar factor of
$1/6$ in $M$-theory, played an important role.
Some additional important details, such as a gravitational correction
to \pxol, were also described there.)
There is no harm in this, because in any event, $M$-theory on $X\times \S^4$
requires a spin structure on $X$.
If $N$ is even or a spin structure is picked on $X$, then quantization
is carried out rather as in section 3.2 by introducing a Heisenberg
group associated with $H^3(M,\Z_N)$.  This Heisenberg group contains
an operator $\Phi_v$ for each $v\in H^3(M,\Z_N)$, with\foot{From one
point of view, the
need for a spin structure arises because the following formula
does not quite determine the Heisenberg group.  To have a Heisenberg
group, we need a multiplication law $\Phi_v\Phi_w=c(v,w)\Phi_{v+w}$,
with $c(v,w)c(w,v)^{-1}=\exp(2\pi i v\cdot w/N)$.  Since $v\cdot w$ is
only well-defined modulo $N$, there is
no elementary formula for $c(v,w)$ except to take an ambiguous
square root $c(v,w)=\pm\sqrt{\exp(2\pi i v\cdot w/N)}$.  A spin
structure determines a consistent set of square roots.
In  fact, having  a Heisenberg group extension
and not just a commutation relation is equivalent -- roughly as we saw
in section 3.3 -- to having a suitable line bundle over the classical
phase space.  As analyzed in detail in \witto, defining this line bundle
depends on having a spin structure on $M$.  The Chern-Simons theory
that we have studied in this paper in 
five dimensions likewise needs a spin structure, as we saw in section
3.3.}
\eqn\tomoko{\Phi_v\Phi_w=\Phi_w\Phi_v\exp\left(2\pi i v\cdot w/N\right).}
We thus have a group extension
\eqn\retur{0\to \Z_N\to W\to H^3(M,\Z_N)\to 0,}
and the quantum Hilbert space $\H$ is an irreducible representation of
$W$.  (Irreducibility can be proved by a detailed analysis along
the lines of sections 3.4-6.)

One important difference from the four-dimensional case is that $W$
is defined using only one copy of $H^3(M,\Z_N)$ (rather than two
copies of $H^2(M,\Z_N)$ as in four dimensions).  So there is no way
to pick a polarization and give an explicit description of $\H$
without using some detailed properties of $W$ and breaking the symmetries
of the finite group $H^3(M,\Z_N)$.  One simple case is $M=\S^1\times Y$,
with $Y$ a five-dimensional spin manifold.  In this case,
$H^3(M,\Z_N)=H^2(Y,\Z_N)\oplus H^3(Y,\Z_N)$.  A polarization of $H^3(M,\Z_N)$
is given by its subgroup $H^3(Y,\Z_N)$.  The quotient is
$H^3(M,\Z_N)/H^3(Y,\Z_N)=H^2(Y,\Z_N)$.  Hence, the Hilbert space
$\H$ has a basis with a vector $\Psi_w$ for each $w\in H^2(Y,\Z_N)$.
This result has a simple explanation.  The $(0,2)$ conformal field
theory, if compactified 
on $\S^1$, gives a theory that looks at low energies like
$SU(N)/\Z_N$ gauge theory in five dimensions.  If this is further
compactified on the five-manifold $Y$, the gauge theory has
a discrete magnetic flux taking values in $H^2(Y,\Z_N)$, and this
leads to the Hilbert space that we found.

\nref\maldacs{J. Maldacena, ``Wilson Loops In Large $N$ Field Theories,''
hep-th/9803002.}
\nref\reyyee{S.-J. Rey and J. Yee, ``Macroscopic Strings As Heavy Quarks
In Large $N$ Gauge Theories,'' hep-th/9803001.}

\newsec{'t Hooft And Wilson Lines In Four Dimensions}

In this concluding section, we will analyze the behavior of 
the $\N=4$ super Yang-Mills theory on a four-manifold $M$ with
't Hooft and Wilson loops included.

We begin on a five-manifold $X$ by supplementing the Chern-Simons
Lagrangian $L_{CS} $ by couplings of the $B$-fields to string worldsheets.
The strings in question are fundamental strings coupling to $B_{NS}$
and $D$-strings coupling to $B_{RR}$.
We take the $D$-string and elementary string worldsheets to be closed
surfaces $S_{RR}$ and $S_{NS}$, which are not necessarily connected.  
Including the couplings to the strings,
the Lagrangian is
\eqn\isno{\widehat L=NI_{CS}(B_{RR},B_{NS})-{i\over 2\pi}\int_{S_{RR}}B_{RR}
-{i\over 2\pi}\int_{S_{NS}}B_{NS}.}
 If $X$ has conformal boundary $M$, then the boundary
of $S_{RR}$ and $S_{NS}$ should be \refs{\maldacs,\reyyee}
one-manifolds $C_{RR} $ and $C_{NS}$ in $M$ 
on which 't Hooft and Wilson loops are inserted.  (The $C$'s are not
necessarily connected.)

The path integral on $X$ will take values in a Hilbert space associated
with the boundary.  Near the boundary, we have $X=M\times \R$, where
we view $\R$ as the ``time'' direction.  Near the boundary, 
we can take $S_{RR}=C_{RR}\times \R$
and $S_{NS}=C_{NS}\times \R$.  Quantization is carried out formally
as in section 3.1.  The Gauss law constraint contains an extra
contribution from the strings and reads
\eqn\osno{\eqalign{ NH_{NS}&+\delta(C_{RR})=0\cr
-NH_{RR}&+\delta(C_{NS})  = 0 . \cr}}
Here, for $C=C_{RR}$ or $C_{NS}$, $\delta (C)$ is a delta-function supported
on $C$, representing  a cohomology class in $H^3(M,\Z)$ that is dual
to $C$.  We will call this class $[C]$.

If the equations \osno\ have solutions, then as these equations are
linear, by shifting the $B$-fields by any solution, we can reduce
to the previous case in which the phase space that must be quantized
is governed by $H_{RR}=H_{NS}=0$.  
Hence, the quantum Hilbert space
is isomorphic to the one obtained without the Wilson lines, though
not canonically so as some arbitrary choice of a solution of
Gauss's law has to be made
in identifying the phase spaces with and without the Wilson lines.
However, the condition that equations \osno\ should have any solutions
at all is nontrivial.  The requirement that these equations have a solution
for some $B$-fields is simply that the elements $[C_{RR}]$, $[C_{NS}]$ should
be divisible by $N$ in $H^3(M,\Z)$.  (When torsion is present, a full
justification
 of this statement requires a torsion extension of the Gauss's law
constraint, along lines presented in section 3.5.)

For example, if $M=Y\times \S^1$, with $Y$ a three-manifold, then the
requirement is that the $C$'s should wrap around $\S^1$ a number
of times divisible by $N$.  
This statement has a simple intuitive interpretation.
It means that the Wilson line wrapped on $C_{NS}$ (or
the magnetic counterpart wrapped on $C_{RR}$) is
invariant under the thermal symmetry group $F=H^1(M,\Z_N)$.
This way of seeing the thermal symmetry
group is actually a close cousin of the argument in \ahaw.

For the rest of this section, we specialize to the case $M=\S^4$.
In this case, since $H^3(M,\Z)=0$,
there is no topological obstruction to solving the Gauss
law constraints, regardless of what the $C$'s might be.  
Moreover, the space of solutions mod gauge transformations is a single
point (since the cohomology groups that classify flat $B$-fields are
all zero).  So the quantum Hilbert space is one-dimensional.

The expectation value of an arbitrary product of Wilson and 't Hooft
loops on $\S^4$ thus takes values in a one-dimensional Hilbert space $\H$.
However, the expectation value cannot in general be understood as
an ordinary complex number, because there is no natural way to pick
a basis vector in $\H$ and identify $\H$ with the complex numbers
${\bf C}$.  Let us explore the matter somewhat more
thoroughly and see what is involved in picking a basis vector.

The presence of the Wilson loop on $C_{NS}$ leads to the presence
in the path integral on $X\times \S^5$ of a factor
\eqn\turmigo{\exp\left(i\int_{S_{NS}}B_{NS}\right).}
This factor is well-defined as a complex number if $S_{NS}$ is a closed
surface.  If not, the factor takes values in a one-dimensional
Hilbert space $\H_{NS}$.  $\H$ is the tensor product of $\H_{NS}$ with
a similar one-dimensional Hilbert space ${\cal H}_{RR}$ associated
with the $D$-strings, as well as a factor independent of all 't Hooft
and Wilson loops that enters in defining the bulk term in the action.
 To trivialize $\H_{NS}$, we should give
a surface $D_{NS}\subset M$ of boundary $-C_{NS}$ (the minus sign
is a reversal of orientation), and replace \turmigo\ by
\eqn\obermigo{\exp\left(i\int_{S_{NS}+D_{NS}}B_{NS}\right).}
Here $S_{NS}+D_{NS}$ is a closed surface in $X$, so \obermigo\ is 
well-defined as a number.\foot{$D_{NS}$ should lie in $M$,
not just in $X$,
so as to trivialize ${\cal H}$ (which only depends on $M$) once and for
all, independent of the choice of $X$.}   
We must, however, investigate the dependence on the choice of $D_{NS}$.
Let $D_{NS}'$ be another choice.  Then in replacing $D_{NS}$ by $D'_{NS}$,
\obermigo\ is multipled by
\eqn\ermigo{\Psi=\exp\left(i\int_E B_{NS}\right),}
where $E$ is the closed surface in $M$ defined by $E=D'_{NS}-D_{NS}$.
Since $E$ is contained in  $\S^4$, whose homology vanishes,
it is the boundary of a three-manifold 
$Y\subset M$, and we can write
\eqn\urmigo{\Psi=\exp\left(i\int_Y H_{NS}\right).}
If there are no $D$-strings,
then $H_{NS}$ vanishes by the equations of motion, and
the factor $\Psi$ in
the path integral can be eliminated by a redefinition of the fields,
in fact, by $B_{RR}\to B_{RR}+(2\pi/N)\delta(Y)$.  ($Y$ is of codimension
two in $X$, and $\delta(Y)$ is  a two-form dual to $Y$.)  But if there
are $D$-strings, this is not quite a symmetry. Under this
transformation, the $D$-string factor $\exp(i\int_{S_{RR}}B_{RR})$ in the path
integral picks up a phase $\exp\left((2\pi i/N) Y\cdot C_{RR}\right)$,
where $Y\cdot C_{RR}$ denotes the oriented intersection number of $Y$
and $C_{RR}$.  That intersection number is (by definition)
the linking number $\ell(E,C_{RR})$.  Hence, under a change in trivialization
of $\H_{NS}$ coming from a change in $D_{NS}$, the path integral
is multiplied by
\eqn\ksonsp{\exp(2\pi i \ell(E,C_{RR})/N).}

This factor is, of course, 1 if there are no 't Hooft loops ($C_{RR}=0$).
So the expectation value of a product of Wilson loops only
(or likewise of 't Hooft loops only) can be viewed as an ordinary number.
But 
the expectation value of a product consisting of both Wilson
and 't Hooft loops, though well-defined as a vector in the one-dimensional
Hilbert space $\H$, is ambiguous up to multiplication by an $N^{th}$
root of unity if one wishes to interpret it as an ordinary complex number.

An ambiguity of just this nature can be seen by standard gauge theoretic
methods.  In that context, 
one begins with the problem of what it means in $SU(N)/\Z_N$ gauge theory
to define the expectation value of a Wilson loop, wrapped on a circle 
$C_{NS}$, in the fundamental
representation of $SU(N)$.
For this, we would want
 a lifting of the $SU(N)/\Z_N$ bundle to $SU(N)$ at least
along $C_{NS}$.  It suffices to give a surface $D_{NS}$ with boundary $C_{NS}$.
For, as $H^2(D_{NS},\Z_N)=0$, the $SU(N)/\Z_N$ bundle when restricted
to $D_{NS}$ can be lifted to $SU(N)$, and though the lifting is not unique,
the nonuniqueness does not affect the value of a Wilson loop that wraps
around the boundary of $D_{NS}$.
The answer one gets this way does depend on the choice of $D_{NS}$ if
't Hooft loops are present also.  Upon changing $D_{NS}$ to
$D'_{NS}$, one gets in the gauge theory, in the presence of an 't Hooft loop
on $C_{RR}$, a change in the Wilson loop expectation value given
by the same formula as in \ksonsp.   That is because, if the linking
number is nonzero, the liftings of the $SU(N)/\Z_N$ bundle to $SU(N)$
on $D_{NS}$ and $D'_{NS}$ disagree on their common boundary.

Of course, different points of view are possible about the
expectation value of a product of Wilson and 't Hooft loops on $\S^4$.
The traditional point of view is to reject such correlation functions on the
grounds that they are not well-defined as complex numbers.
If one chooses to consider such correlation functions,
we have shown how they must be interpreted, and more specifically we have
shown that they have the same meaning whether considered in gauge theory
on $\S^4$ or in string theory on $AdS_5\times \S^5$.

\bigskip
This work was supported in part by NSF Grant PHY-9513835.  I would
like to thank P. Deligne,
D. Freed, O. Ganor, M. Hopkins, N. Seiberg, and S. Sethi for discussions.

\listrefs
\end